\documentclass[
    ,final            
  ]
  {aipproc}

\layoutstyle{6x9}

\newcommand{\ba}{\begin{eqnarray}}
\newcommand{\ea}{\end{eqnarray}}

\begin{document}

\title{Partial Dynamical Symmetry and Anharmonicity in Gamma-Soft Nuclei}


\author{A. Leviatan}{
  address={
Racah Institute of Physics, The Hebrew University, 
Jerusalem 91904, Israel}
}

\begin{abstract}
Partial dynamical symmetry is shown to be relevant for describing the 
anharmonicity of excited bands in $^{196}$Pt while retaining 
solvability and good $SO(6)$ symmetry for the ground band.
\end{abstract}

\maketitle


Gamma-soft nuclei can be described in the interacting boson model (IBM) 
in its SO(6) dynamical symmetry (DS) limit~\cite{Iachello87}. 
The latter limit corresponds to the chain of nested algebras 
\begin{equation}
\begin{array}{ccccccccc}
{\rm U}(6)&\supset&{\rm SO}(6)&\supset&{\rm SO}(5)&
\supset&{\rm SO}(3)&\supset&{\rm SO}(2)\\
\downarrow&&\downarrow&&\downarrow&&\downarrow&&
\downarrow\\
[0mm]
[N]&&\langle\Sigma\rangle&&(\tau)&\nu_\Delta& L&&M
\end{array}
\label{chainso6}
\end{equation}
where, below each algebra, its associated labels of irreducible 
representations (irreps) are given 
and $\nu_\Delta$ is a multiplicity label. 
The eigenstates $|[N]\langle\Sigma\rangle(\tau)\nu_\Delta LM\rangle$
are obtained with a Hamiltonian
with SO(6) DS which can be transcribed in the form
\ba
\hat{H}_{\rm DS} &=& A\,\hat{P}_{+}\hat{P}_{-} +
B\, \hat{C}_{{\rm SO(5)}} + C\,\hat{C}_{{\rm SO(3)}}.
\label{hDS}
\ea
Here $\hat{C}_{G}$ denotes the quadratic Casimir operator of $G$,
$\hat{P}_{+}\equiv{1\over2}(s^\dag s^\dag-d^\dag\cdot d^\dag)$,
$4\hat{P}_{+}\hat{P}_{-} = \hat{N}(\hat{N}+4)-\hat C_{{\rm SO(6)}}$
and $\hat{P}_{-}= \hat{P}_{+}^{\dagger}$. The monopole $(s)$ 
and quadrupole $(d)$ bosons represent valence nucleon pairs 
whose total number, $\hat{N}=\hat{n}_s + \hat{n}_d$, is conserved. 
The SO(6)-DS Hamiltonian, $\hat{H}_{\rm DS}$, is completely solvable 
with eigenenergies
\ba
E_{\rm DS} &=& {\textstyle{\frac{1}{4}}}
A\,(N-\Sigma)(N+\Sigma + 4) + B\,\tau(\tau+3)+\, C\,L(L+1).
\label{eDS}
\ea
\begin{figure}
  \includegraphics[height=.28\textheight]{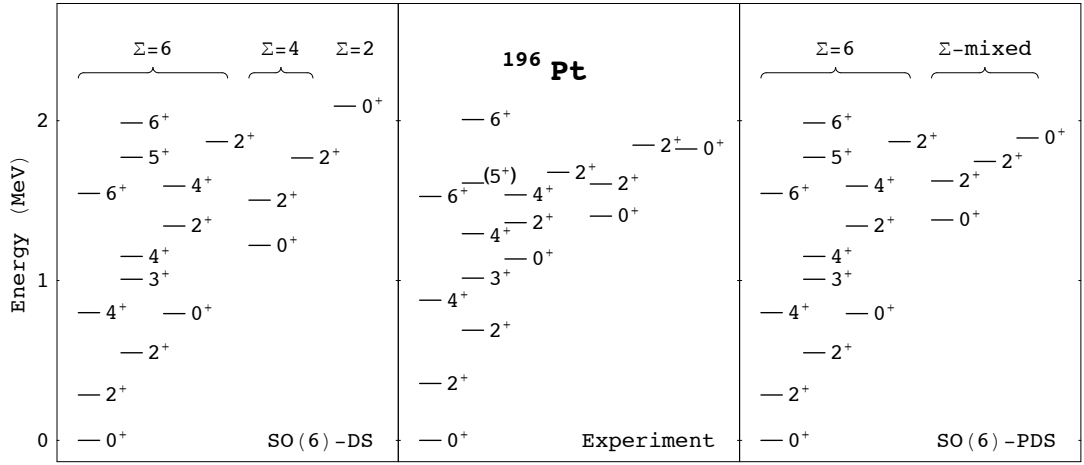}
\caption{
Observed spectrum of $^{196}$Pt~\cite{NDS07}
compared with the calculated spectra
of $\hat H_{\rm DS}$ (\ref{hDS}),
with SO(6) dynamical symmetry (DS),
and of $\hat H_{\rm PDS}$ (\ref{hPDS}) with partial dynamical symmetry (PDS).
The parameters in $\hat H_{\rm DS}$ $(\hat H_{\rm PDS})$ are
$A=174.2\, (122.9)$,
$B=44.0\, (44.0)$,
$C=17.9\, (17.9)$,
and $\eta=0\, (34.9)$ keV.
The boson number is $N=6$
and $\Sigma$ is an SO(6) label. From~\cite{ramos09}.}
\label{pt196}
\end{figure}
\begin{table}
\begin{tabular}{clll|cllll}
\hline
\tablehead{1}{c}{b}{Transition} & 
\tablehead{1}{l}{b}{Experiment} & 
\tablehead{1}{l}{b}{DS} & 
\tablehead{1}{l}{b}{PDS} \hspace{0.5cm} &  
\tablehead{1}{c}{b}{Transition} & 
\tablehead{1}{l}{b}{Experiment} & 
\tablehead{1}{l}{b}{DS} & 
\tablehead{1}{l}{b}{PDS} \\
\hline
$2^+_1\rightarrow0^+_1$& 0.274~(1)  &  0.274&  0.274 & 
$2^+_3\rightarrow0^+_2$& 0.034~(34) &  0.119&  0.119 \\
$2^+_2\rightarrow2^+_1$& 0.368~(9)  &  0.358&  0.358 &
$2^+_3\rightarrow4^+_1$& 0.0009~(8) &  0.0004& 0.0004 \\
$2^+_2\rightarrow0^+_1$& 3.10$^{-8}$(3) & 0.0018& 0.0018 &
$2^+_3\rightarrow2^+_2$& 0.0018~(16)& 0.0013& 0.0013 \\
$4^+_1\rightarrow2^+_1$& 0.405~(6)  &  0.358&  0.358 &
$2^+_3\rightarrow0^+_1$& 0.00002~(2)& 0        & 0   \\
$0^+_2\rightarrow2^+_2$& 0.121~(67) &  0.365&  0.365 &
$6^+_2\rightarrow6^+_1$& 0.108~(34) & 0.103& 0.103   \\
$0^+_2\rightarrow2^+_1$& 0.019~(10) &  0.003&  0.003 &
$6^+_2\rightarrow4^+_2$& 0.331~(88) &  0.221&  0.221 \\
$4^+_2\rightarrow4^+_1$& 0.115~(40) &  0.174&  0.174 &
$6^+_2\rightarrow4^+_1$& 0.0032~(9) & 0.0008& 0.0008 \\
$4^+_2\rightarrow2^+_2$& 0.196~(42) &  0.191&  0.191 &
$0^+_3\rightarrow2^+_2$&$<0.0028$   & 0.0037& 0.0028 \\
$4^+_2\rightarrow2^+_1$& 0.004~(1)  &  0.001&  0.001 &
$0^+_3\rightarrow2^+_1$&$<0.034$    & 0          & 0 \\
$6^+_1\rightarrow4^+_1$& 0.493~(32) &  0.365&  0.365 &
                       &            &                & \\
\hline
\end{tabular}
\caption{
Observed~\cite{NDS07} and calculated $B$(E2) values
(in $e^2{\rm b}^2$) for $^{196}$Pt.
For both the exact (DS) and partial (PDS)
SO(6) dynamical symmetry calculations, the E2 operator is
$e_{\rm b}[(s^\dag\times\tilde d+d^\dag\times\tilde s)^{(2)}
+\chi(d^\dag\times\tilde d)^{(2)}]$ with
$e_{\rm b}=0.151$ $e$b and $\chi=0.29$. From~\cite{ramos09}.}
\label{be2}
\end{table}
The spectrum resembles that of a $\gamma$-unstable deformed rotor,
where states are arranged in bands with SO(6) quantum number
$\Sigma=N-2v$, $(v=0,1,2,\ldots)$.
The in-band rotational structure is governed by the SO(5) and 
SO(3) terms in $\hat{H}_{\rm DS}$ (\ref{hDS}), 
with characteristic $\tau(\tau+3)$ and $L(L+1)$ splitting.
A comparison with the experimental 
spectrum and E2 rates of $^{196}$Pt~\cite{NDS07} 
is shown in Fig.~\ref{pt196} and Table~\ref{be2}. 
It displays a good description
for properties of states in the ground band $(\Sigma=N)$.
This observation was the basis of the claim~\cite{Cizewski78}
that the SO(6)-DS is manifested empirically in $^{196}$Pt.
However, the resulting fit to energies of excited bands is quite poor.
The $0^+_1$, $0^+_3$, and $0^+_4$ levels of $^{196}$Pt
at excitation energies 0, 1403, 1823 keV, respectively,
are identified as the bandhead states
of the ground $(v=0)$, first- $(v=1)$
and second- $(v=2)$ excited vibrational bands~\cite{Cizewski78}.
Their empirical anharmonicity,
defined by the ratio $R=E(v=2)/E(v=1)-2$,
is found to be $R=-0.70$.
In the SO(6)-DS limit these bandhead states
have $\tau=L=0$ and $\Sigma=N,N-2,N-4$, respectively.
The anharmonicity $R=-2/(N+1)$,
as calculated from Eq.~(\ref{eDS}), is fixed by $N$.
For $N=6$, which is the appropriate boson number for $^{196}$Pt,
the SO(6)-DS value is $R=-0.29$,
which is in marked disagreement with the empirical value.
A detailed study of double-phonon excitations within the IBM,
has concluded that large anharmonicities can be incorporated
only by the inclusion of at least cubic terms in the
Hamiltonian~\cite{ramos00b}.
In the IBM there are 17 possible three-body interactions.
One is thus confronted with the need to select suitable higher-order terms
that can break the DS in excited bands but preserve it in the ground band.
These are precisely the defining 
properties of a partial dynamical symmetry (PDS). 
The essential idea 
is to relax the stringent conditions of {\em complete} solvability, 
so that only part of the eigenspectrum retains all the DS quantum numbers.  
Various types of PDS are known to be relevant to nuclear 
spectroscopy~[5-11], to systems with mixed chaotic and
regular dynamics~\cite{WAL93,LW96} and to
quantum phase transitions~\cite{lev07}.
In the present contribution we demonstrate the relevance of PDS to 
the anharmonicity of excited bands in $^{196}$Pt~\cite{ramos09}.
\begin{table}
\begin{tabular}{lrrrr}
\hline
\tablehead{1}{c}{b}{Bandhead}
  & \tablehead{1}{c}{b}{$\Sigma=6$} 
  & \tablehead{1}{c}{b}{$\Sigma=4$} 
  & \tablehead{1}{c}{b}{$\Sigma=2$} 
  & \tablehead{1}{c}{b}{$\Sigma=0$}\\
\hline
$0^{+}(v=0)$ \hspace{0.1cm}
 & 100 \% \hspace{0.1cm} &        &         &        \\
$0^{+}(v=1)$  &         & 76.5 \% \hspace{0.1cm} & 16.1 \%  
\hspace{0.1cm} & 7.4 \%  \\
$0^{+}(v=2)$  &         & 19.6 \% \hspace{0.1cm} & 18.4 \% 
\hspace{0.1cm} & 62.0 \% \\
\hline
\end{tabular}
\caption{$SO(6)$ decomposition of eigenstates of 
$\hat{H}_{\rm PDS}$~(\ref{hPDS}), corresponding to bandhead states in 
$^{196}$Pt.}
\label{tab2}
\end{table}

Hamiltonians with SO(6) PDS preserve the analyticity of only a 
{\em subset} of the states~(\ref{chainso6}).
The construction of interactions with this property
requires $n$-boson creation and annihilation operators, 
$\hat B^\dag_{[n]\langle\sigma\rangle(\tau)\ell m}$ and
$\tilde{B}_{[n^5]\langle\sigma\rangle(\tau)\ell m}$, 
with definite tensor character in the basis~(\ref{chainso6}). 
Of particular interest are $n$-boson annihilation operators which satisfy 
\begin{equation}
\tilde{B}_{[n^5]\langle\sigma\rangle(\tau)\ell m}
|[N]\langle N\rangle(\tau)\nu_\Delta LM\rangle=0,
\label{anniso6}
\end{equation}
for all possible values of $\tau,L$ contained in the
SO(6) irrep $\langle N\rangle$.
The annihilation condition~(\ref{anniso6}) is satisfied for 
tensor operators with $\sigma<n$.
This is so because the action of
$\tilde{B}_{[n^5]\langle\sigma\rangle(\tau)\ell m}$
leads to an $(N-n)$-boson state
that contains the SO(6) irreps
$\langle\Sigma\rangle=\langle N-n-2i\rangle,\,i=0,1,\dots$, 
which cannot be coupled with $\langle\sigma\rangle$
to yield $\langle\Sigma\rangle=\langle N\rangle$, since $\sigma<n$.
Number-conserving normal-ordered interactions that are constructed out
of such tensors (and their Hermitian conjugates) 
thus have $|[N]\langle N\rangle(\tau)\nu_\Delta LM\rangle$
as eigenstates with zero eigenvalue.

A systematic enumeration of all interactions with this property
is a simple matter of SO(6) coupling. For example, SO(6) tensors, 
$\hat B^\dag_{[n]\langle\sigma\rangle(\tau)\ell m}$,
with $\sigma < n=2$ or $\sigma < n=3$ are found to be
\begin{equation}
\hat B^\dag_{[2]\langle0\rangle(0)00} \propto \hat P_+,
\quad
\hat B^\dag_{[3]\langle1\rangle(1)2m} \propto
\hat P_+d^\dag_m,
\quad
\hat B^\dag_{[3]\langle1\rangle(0)00} \propto
\hat P_+s^\dag~.
\label{twothree}
\end{equation}
The two-boson SO(6) tensor gives rise to a two-body   
SO(6)-invariant interaction, $\hat P_+\hat P_-$, 
which is simply the completely solvable 
SO(6) term in $\hat{H}_{\rm DS}$, Eq.~(\ref{hDS}). 
From the three-boson SO(6) tensors one can construct three-body 
interactions with an SO(6) PDS, namely, 
$\hat P_+\hat n_s\hat P_-$
and $\hat P_+\hat n_d\hat P_-$. 
Since the combination $\hat P_+(\hat n_s+\hat n_d)\hat P_-
= (\hat{N} -2)\hat{P}_{+}\hat{P}_{-}$
is completely solvable in SO(6),
there is only one genuine partially solvable three-body interaction
which can be chosen as $\hat P_+\hat n_s\hat P_-$,
with tensorial components $\sigma=0,2$.

On the basis of the preceding discussion we propose 
to use the following Hamiltonian with SO(6)-PDS
\begin{equation}
\hat{H}_{\rm PDS}=\hat{H}_{\rm DS}+\eta\hat{P}_{+}\hat{n}_s\hat{P}_{-},
\label{hPDS}
\end{equation}
where the terms are defined in Eqs.~(\ref{hDS}) and (\ref{twothree}).
The spectrum of $\hat{H}_{\rm PDS}$ is shown in Fig.~\ref{pt196}.
The states belonging to the $\Sigma=N=6$ multiplet remain solvable
with energies given by the same DS expression, Eq.~(\ref{eDS}).
As shown in Table~\ref{tab2}, 
states with $\Sigma < 6$ are generally admixed 
but agree better with the data than in the DS calculation. 
Thus, although the ground band is pure,
the excited bands exhibit strong SO(6) breaking.
The calculated SO(6)-PDS anharmonicity for these bands is $R=-0.63$,
much closer to the empirical value, $R=-0.70$.
We emphasize that not only the energies
but also the wave functions of the $\Sigma=N$ states remain unchanged
when the Hamiltonian is generalized from DS to PDS.
Consequently, the E2 rates for transitions among this class of states
are the same in the DS and PDS calculations.
This is evident in Table~\ref{be2}
where most of the E2 data concern transitions between $\Sigma=N=6$ states.
Only transitions involving states from excited bands ({\it e.g.},
the $0^+_3$ state in Table~\ref{be2}) can distinguish between DS and PDS.

A similar procedure can be implemented on a general 
dynamical symmetry chain 
\begin{equation}
\begin{array}{ccccccc}
G_{\rm dyn}&\supset&G&\supset&\cdots&\supset&G_{\rm sym}\\
\downarrow&&\downarrow&&&&\downarrow\\[0mm]
[h_N]&&\langle\Sigma\rangle&&&&\Lambda
\end{array}
\label{chain}
\end{equation}
where $G_{\rm dyn}$ and $G_{\rm sym}$ are, respectively, the dynamical 
and symmetry algebras of the system. For $N$ identical particles
the irrep $[h_N]$ is either symmetric $[N]$ (bosons)
or antisymmetric $[1^N]$ (fermions). 
Hamiltonians which preserve the solvability 
of states with $\langle\Sigma\rangle=\langle\Sigma_0\rangle$, 
involve $n$-particle annihilation tensor operators satisfying 
\begin{equation}
\hat T_{[h_n]\langle\sigma\rangle\lambda}
|[h_N]\langle\Sigma_0\rangle\Lambda\rangle=0,
\label{anni}
\end{equation}
for all possible values of $\Lambda$
contained in the given $G$-irrep~$\langle\Sigma_0\rangle$.
The solution of condition~(\ref{anni}) amounts to carrying out a $G$ 
Kronecker product $\langle\sigma\rangle\times\langle\Sigma_0\rangle$. 
This establishes a 
generic and systematic procedure 
for identifying and selecting interactions, 
of a given order, with PDS. The resulting Hamiltonians break the DS but 
retain selected subsets of solvable eigenstates with good symmetry.
As demonstrated in the present contribution, 
the advantage of using higher-order interactions with PDS
is that they can be introduced without destroying results
previously obtained with a DS for a segment of the spectrum.

This contribution is based on work done in collaboration 
with J.E.~Garc\'\i a-Ramos (Huelva) and P. Van Isacker (GANIL) 
and is supported by grants from the ISF and BSF.

\end{document}